\documentclass[10pt]{article}
\usepackage[utf8]{inputenc}
\usepackage{amsmath}
\usepackage{amssymb}
\usepackage{geometry}
\usepackage[superscript]{cite}
\usepackage{graphicx}
\usepackage{caption}
\usepackage{subcaption}
\usepackage{setspace}
\usepackage{lipsum}
\usepackage{mathtools}
\usepackage{cuted}
\usepackage[english]{babel}
\usepackage{hyperref}
\usepackage{multicol}
\usepackage[dvipsnames]{xcolor}
\usepackage{multirow}
\title{\textbf{Volume Expansive Pressure (VEP) Driven Non-Trivial Topological Phase Transition In LiMgBi}}
\author{Raghottam M Sattigeri$^{1,a}$, Sharad Babu Pillai$^{1,b}$, Prafulla K Jha$^{1,c}$ and Brahmananda Chakraborty$^{2,d}$}
\date{\small$^{1}$\textit{Department of Physics, Faculty of Science, The Maharaja Sayajirao University of Baroda, Vadodara.\\ $^{2}$High Pressure and Synchrotron Radiation Physics Division, Bhabha Atomic Research Centre, Mumbai.}\\ \footnotesize Corresponding Authors: $^{a}$raghottam.ms@gmail.com, $^{b}$sbpillai001@gmail.com, $^{c}$prafullaj@yahoo.com, $^{d}$brahma@barc.gov.in}

\begin{document}

\maketitle
\doublespacing

\begin{abstract}
\doublespacing
Topological Insulators (TI) exhibit robust spin-locked dissipationless Fermion transport along the surface states. In the current study, we use \textit{first-principles} calculations to investigate a Topological Phase Transition (TPT) in a Half-Heusler (HH) compound LiMgBi driven by a Volume Expansive Pressure (VEP) which is attributed to the presence of, intrinsic voids, thermal perturbations and/or due to a phenomena known as cavity nuclei. We find that, the dynamically stable \textit{face-centred cubic} (FCC) structure of LiMgBi (which belongs to the F$\overline{4}$3m[216] space group), undergoes TPT beyond a critical VEP (at 4.0\%). The continuous application of VEP from 0.0\% to 8.0\% results in a phase transition from a, band insulator to a Dirac semi-metal nature. Qualitatively, the Dirac cone formation and band inversion along the high symmetry point $\mathbf{\Gamma}$ in the Brillouin Zone (BZ) are analysed in terms of Electronic Band Structure (EBS) and Projected Local Density of States (LDOS). The TPT is further characterised by the $\mathbb{Z}_2$ invariant, \big($\nu_0$, $\nu_1$ $\nu_2$ $\nu_3$\big) $\equiv$ \big(1, 0 0 0\big) along the \big(0001\big) surface which indicates quantitatively that, HH LiMgBi is a \textit{strong} TI. We hence propose, HH LiMgBi (known for its piezoelectric, thermo-electric and semi-conducting applications) as a strong TI with potential multi-purpose application in the field of electronics, spintronics and quantum computation.
\end{abstract}


\subsection*{Introduction}

Topological Insulating (TI) phase of matter is interesting due to the exotic quantum properties such as, spin-locked dissipationless transport of Fermions along the topologically protected surfaces. As a result of this, the spin helicity along the surface has several potential applications in the field of spintronics, quantum computation etc \cite{moore2010birth}. Till date, several three dimensional (3D) materials have been predicted theoretically and some are realised experimentally \cite{fu2007topological,hasan2010colloquium,hasan2011three,qi2011topological,ando2013topological} to exhibit TI nature. A variety of 3D compounds (such as, Bi$_2$Se$_3$, $\beta$-As$_2$Te$_3$, AMgBi (A = Li, Na, K) and elemental tellurium (Te)) exhibit TI nature driven by strain \cite{aramberri2018strain,pal2014strain,agapito2013novel}. These materials also have multi-purpose applications \cite{casper2012half,roy2012half,yadav2015first}. It is known that, specific electronic properties of material depend explicitly on their crystal structure and associated space group. This implies that, crystals of different structural class with previously known applications can be designed precisely to investigate for the Topological Phase Transition (TPT) by employing well known band engineering technique which is based on application of strain or pressure (compressive or expansive in nature). 

Strain engineering of the electronic band structure is utilized to realize a strong TI \cite{aramberri2018strain,pal2014strain}. This is because, the application of strain / pressure enhances the strength of Spin Orbit Coupling (SOC) in some systems \cite{pal2014strain}. But, there are systems such as, 3D elemental Te (three fold screw symmetric helical chains of Te atoms) where, it is observed that, weak van der Waals interaction between the helical layered arrangement of Te atoms can be manipulated by application of shear strain (violates three fold screw symmetry of helices) and hydrostatic pressure leading to band inversion which is a characteristic property of a TI system \cite{agapito2013novel}. This suggests that, without venturing into the computationally costly; SOC calculations, simple application of strain or pressure can yield the desired TPT in materials \cite{agapito2013novel,monserrat2017antiferroelectric}. One such system is, Anti-Ferroelectric TI (AFTI) material AMgBi (A = Li, Na, K) which exhibits strong TI nature characterized by the $\mathbb{Z}_2$ invariant $\nu$ = 1 in both Pnma and P6$_3$mc structures \cite{monserrat2017antiferroelectric}. This is achieved under the influence of epitaxial strain and hydrostatic pressure \cite{monserrat2017antiferroelectric}. Off these, the most promising candidate TI was found to be LiMgBi under epitaxial strain leading to AFTI of Type I (represents normal anti-polar and topological polar states) \cite{monserrat2017antiferroelectric}.

LiMgBi which belongs to Pnma and P6$_3$mc space groups and exhibits AFTI of Type I \cite{monserrat2017antiferroelectric} also exists as a Half Heusler (HH) compound (in the form PQR where, P and Q are transition metals and R is a p-block element) with a variety of applications such as a semi-conducting, piezoelectric and thermo-electric material \cite{mooser1960semiconducting,roy2012half,yadav2015first}. Theoretical search for TI nature in HH compounds has been pursued since a long time \cite{lin2015theoretical}. Concrete efforts have led to theoretical predictions of HH compounds as candidate TI's \cite{xiao2010half,lin2010half,yan2014half,liu2011metallic,wang2014topological,li2013electronic,kaur2017ti,moore2009topological,franz2010topological}. Also, there are studies where, strain tuning led to realization of TI nature in materials \cite{feng2012strain,feng2011three,liu2014manipulating,moore2009topological,franz2010topological}. We perform analogous calculations to investigate TPT in \textit{Face-Centred Cubic} (FCC) HH LiMgBi described by the F$\overline{4}$3m[216] space group in Hermann-Mauguin representation. Experimentally it is observed that, LiMgBi in FCC phase exhibits \textit{trivial} band insulating behaviour \cite{roy2012half}. One study suggests that, when screening potential method is employed \cite{liu2018screening}, a large number of candidate HH compounds can be classified as TI and Normal Insulator (NI). Based on this study it turns out that, thirty three HH compounds exhibit TI nature, but, LiMgBi exhibits NI nature at ambient conditions. Feng et.al. show that, strain tuning of band order in cubic systems can lead to TI nature \cite{feng2012strain}. 

With this background, we investigate the possibility of TPT in FCC LiMgBi by strain tuning the band order which is quite uncharted. We find that, under Volume Expansive Pressure (VEP), HH LiMgBi undergoes a TPT at a critical value of 4.0\% and exhibits a strong TI nature characterized by the $\mathbb{Z}_2$ invariant as \big(1, 0 0 0\big). Thus, indicating that, HH LiMgBi is a candidate TI material for novel applications in the field of spintronics, quantum computation etc.

\subsection*{Methodology}

We performed density functional theory based \textit{first-principles} calculations to investigate TPT under the application of pressure. For this purpose, we obtained crystallographic information for LiMgBi in F$\overline{4}$3m[216] space group from MaterialsProject repository \cite{Jain2013}. Optimization of lattice constant (a) was performed following the convergence test for total energy with respect to, wave function cut off and \textbf{k}-mesh. We used norm conserving pseudopotentials under Generalized Gradient Approximation (GGA) which considers 1s$^1$, 3s$^2$ and 6s$^2$6p$^3$ orbitals of Li, Mg and Bi respectively. The pseudopotential method is based on Martins-Troullier with exchange correlation of Perdew-Burke-Ernzerhof (PBE) functional type \cite{perdew1998perdew}. The optimization was performed by, finding a global minima in terms of the total energy of the system and then narrowing down (using bisection method) to a local minima with a constrain that, the total pressure on the atoms is 0.00 kbar. The converged value of plane wave kinetic energy cut-off and charge density are 50 Ry and 200 Ry respectively. A uniform Monkhorst-Pack Grid \cite{monkhorst1976special} for \textbf{k}-points of 7 $\times$ 7 $\times$ 7 was used in the self-consistent calculations. For better prediction of energy band gap \cite{heyd2004efficient}, we performed calculations with Heyd–Scuseria–Ernzerhof (HSE06) screened Coulomb hybrid functional \cite{heyd2003hybrid} in Vienna \textit{ab initio} simulation package (VASP) \cite{PhysRevB.54.11169} with proper optimization. Phonon calculations were performed using Density Functional Perturbation Theory (DFPT) \cite{RevModPhys.73.515}. For phonon calculations, we used 10 $\times$ 10 $\times$ 10 \textbf{q}-mesh which was followed by plotting the dynamical matrices in the entire Brillouin Zone (BZ). Following the analysis of structural and dynamic stability, we performed electronic calculations pertaining to the Electronic Band Structure (EBS), Density of States (DOS) and Surface States (SS). Optimization and electronic properties were calculated using the QUANTUM ESPRESSO code \cite{QE-2017} while the SS were calculated using WannierTools (WT) \cite{WU2017} package which employed the Wannier Function Centres (WFC) obtained from a post processing package Wannier90 (W90) \cite{MOSTOFI20142309}. The electronic properties led to qualitative results whereas for quantitative characterization of the TI nature, WT package was used for thorough $\mathbb{Z}_2$ analysis.

\subsection*{Results and Discussion}

\subsubsection*{Structural Properties and Dynamic Stability}

\begin{figure}[h]
	\centering
	\includegraphics[width = 4in]{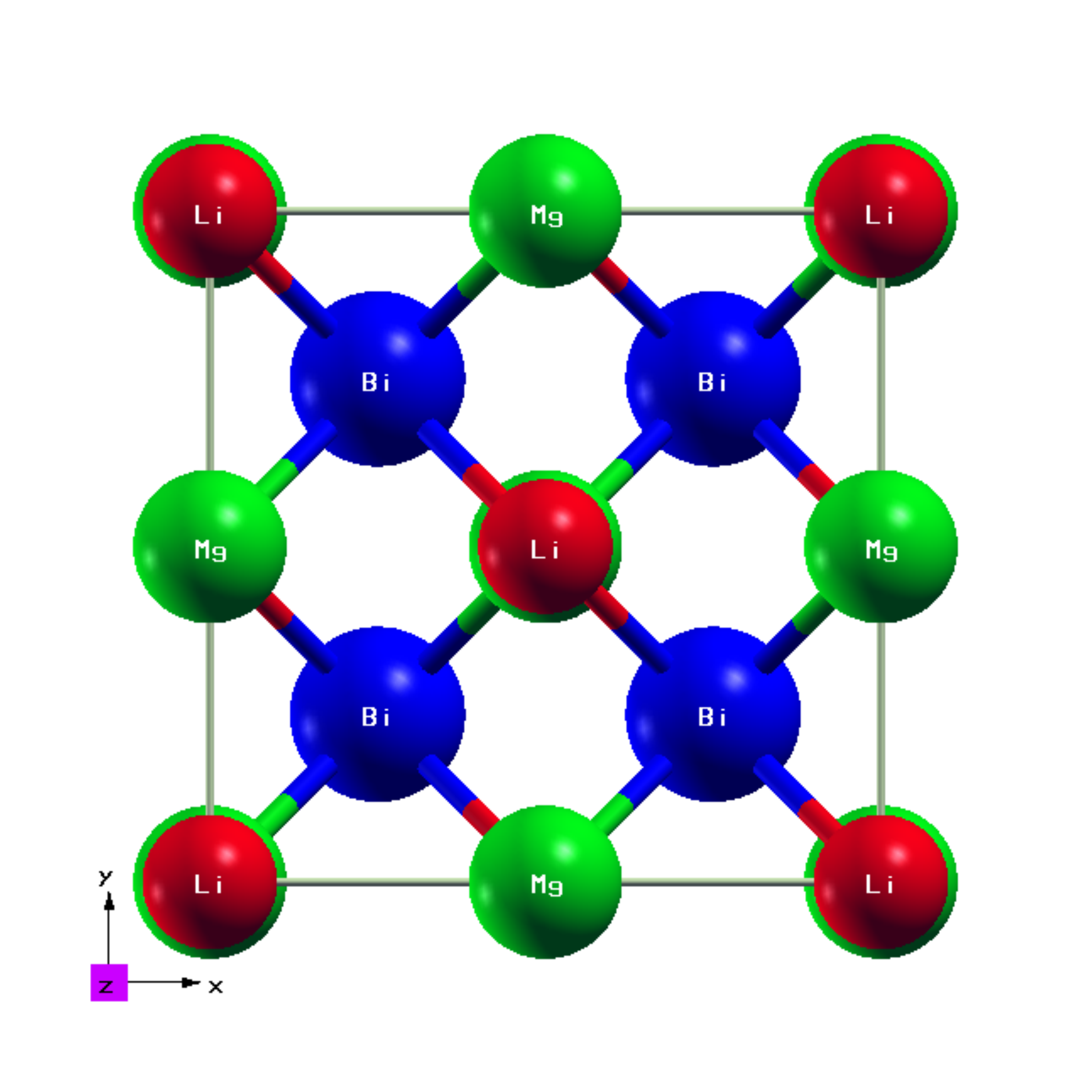}
	\caption{FCC structure of LiMgBi with lattice constant (a) $\sim$ 6.81 \AA}
	\label{fig:fcc}
\end{figure}

\begin{figure}[h]
	\centering
	\includegraphics[width = 4in]{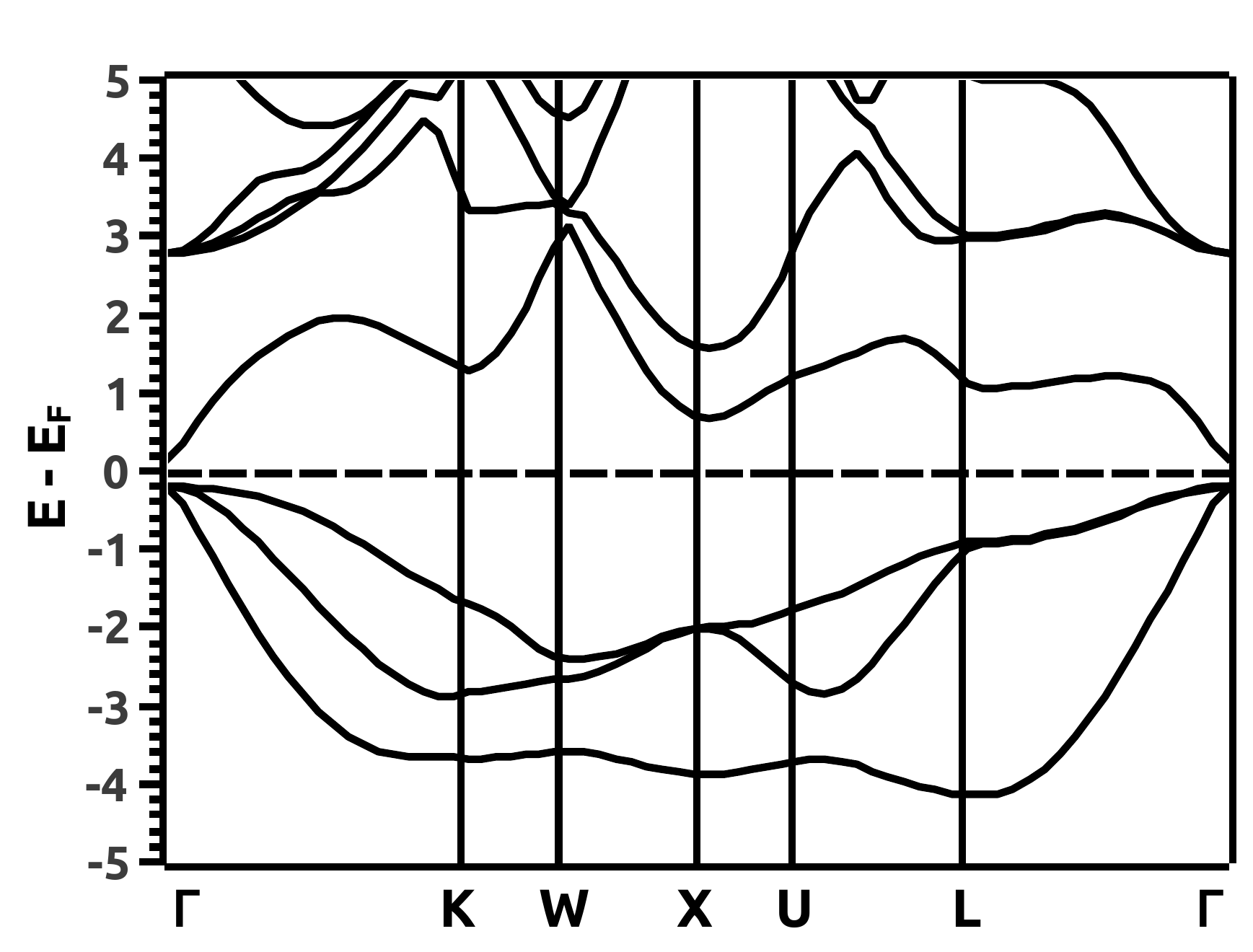}
	\caption{Electronic band structure of HH LiMgBi}
	\label{fig:ebs_tot}
\end{figure}

The optimized value of the lattice constant (a) under GGA approximation is 6.81 \AA. This value is in descent agreement with the experimental (6.74 \AA) and other theoretical (6.71 \AA) values \cite{roy2012half}. The optimized crystal structure is shown in Figure \ref{fig:fcc} which was found to be dynamically stable. The primitive cell vectors for HH LiMgBi are defined in terms of lattice constant `a' as, v$_1$ = (a$/$2)(-1,0,1),  v$_2$ = (a$/$2)(0,1,1), v$_3$ = (a$/$2)(-1,1,0) with atoms Li, Mg and Bi occupying, 4\textit{b}, 4\textit{c} and 4\textit{a} Wyckoff positions respectively \cite{al2010topological}. This structure results in a direct Band Gap (BG) as seen from the EBS presented in Figure \ref{fig:ebs_tot} which is quite small as compared to previous studies (Table \ref{tab1}). Since it is known that, the error in BG prediction is smaller in HSE calculations as compared to pure DFT calculations \cite{heyd2004efficient}, we performed HSE calculations at 0.0\%, 4.0\% and 6.5\% VEP. The BG from HSE calculations are in strong agreement (Table \ref{tab1}) with previous studies and better than the GGA calculations.

\begin{table}[ht]
	\centering
	\begin{tabular}{c c c}
		\hline
			& Work & LiMgBi \\
		\hline\hline
			\multirow{3}{11em}{Lattice Constant (\AA)} & PBE & 6.81 \\ 
			& Theory \cite{roy2012half} & 6.71 \\
			& Experiment \cite{roy2012half,belsky2002new} & 6.74 \\
		\hline\hline
			\multirow{3}{11em}{Energy Band Gap (eV)} & PBE & 0.35 \\ 
			& PBE+HSE & 0.70 \\ 
			& Theory \cite{roy2012half} & 0.62 \\
			& Experiment & - \\
		\hline
	\end{tabular}
	\caption{Theoretical and experimental values of lattice constant (a) and energy band gap.}
	\label{tab1}
\end{table}

\begin{figure}[h]
	\centering
	\includegraphics[width = 4in]{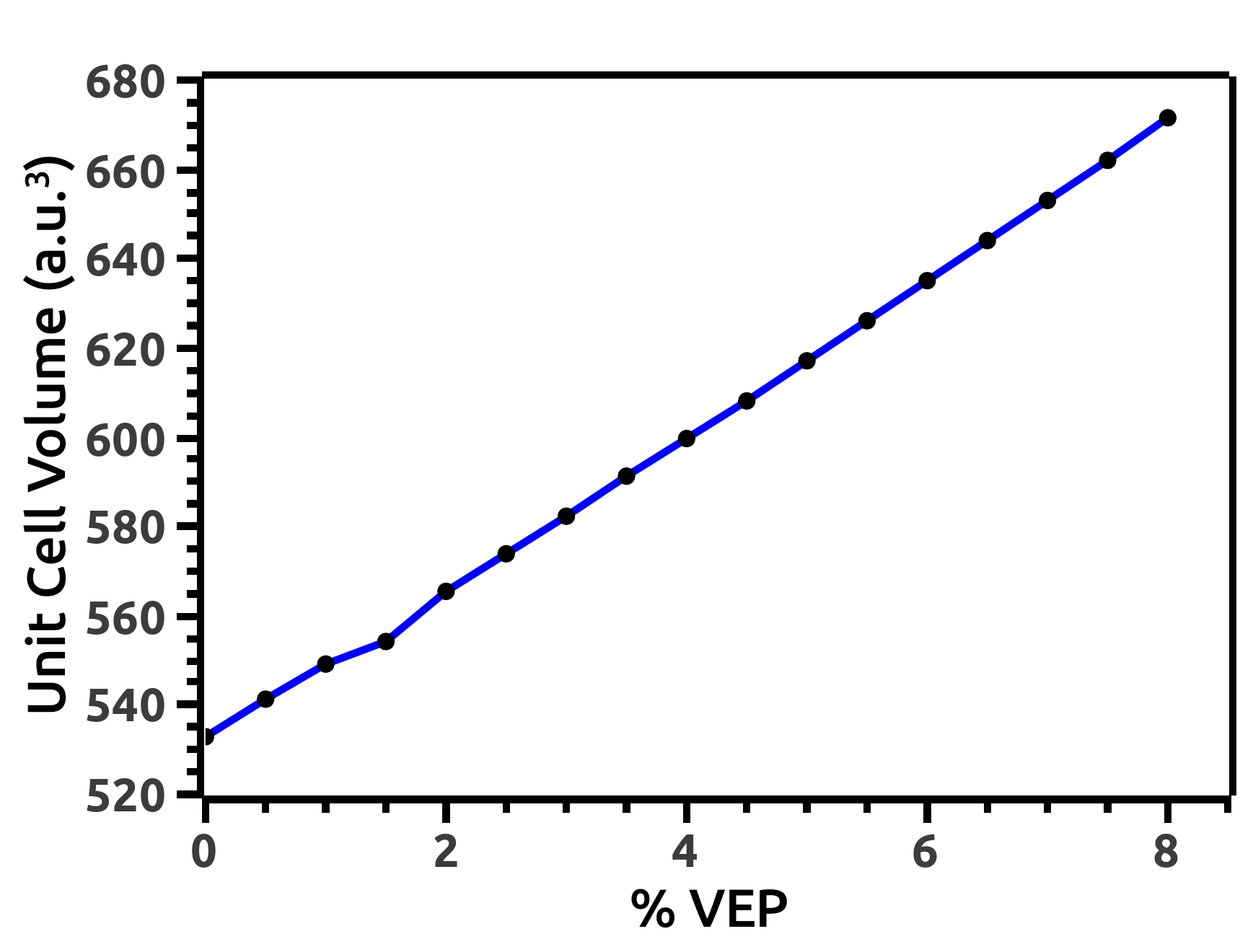}
	\caption{Increment in the unit cell volume (a.u.$^3$) due to VEP}
	\label{fig:unit_vol}
	\includegraphics[width = 4in]{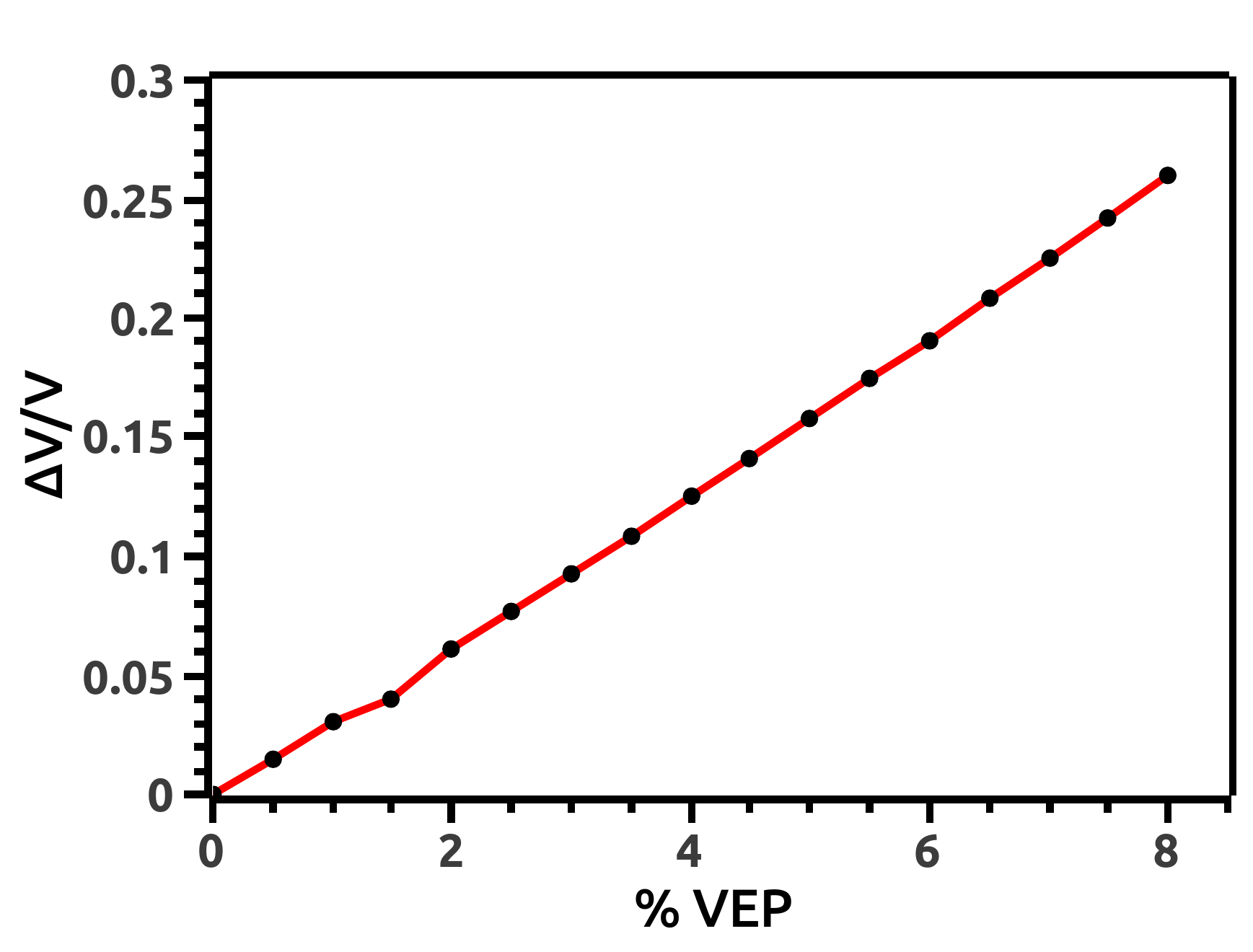}
	\caption{Volume strain with respect to \% VEP}
	\label{fig:vol_str}
\end{figure}

\begin{figure}[h]
	\centering
	\includegraphics[width = 4in]{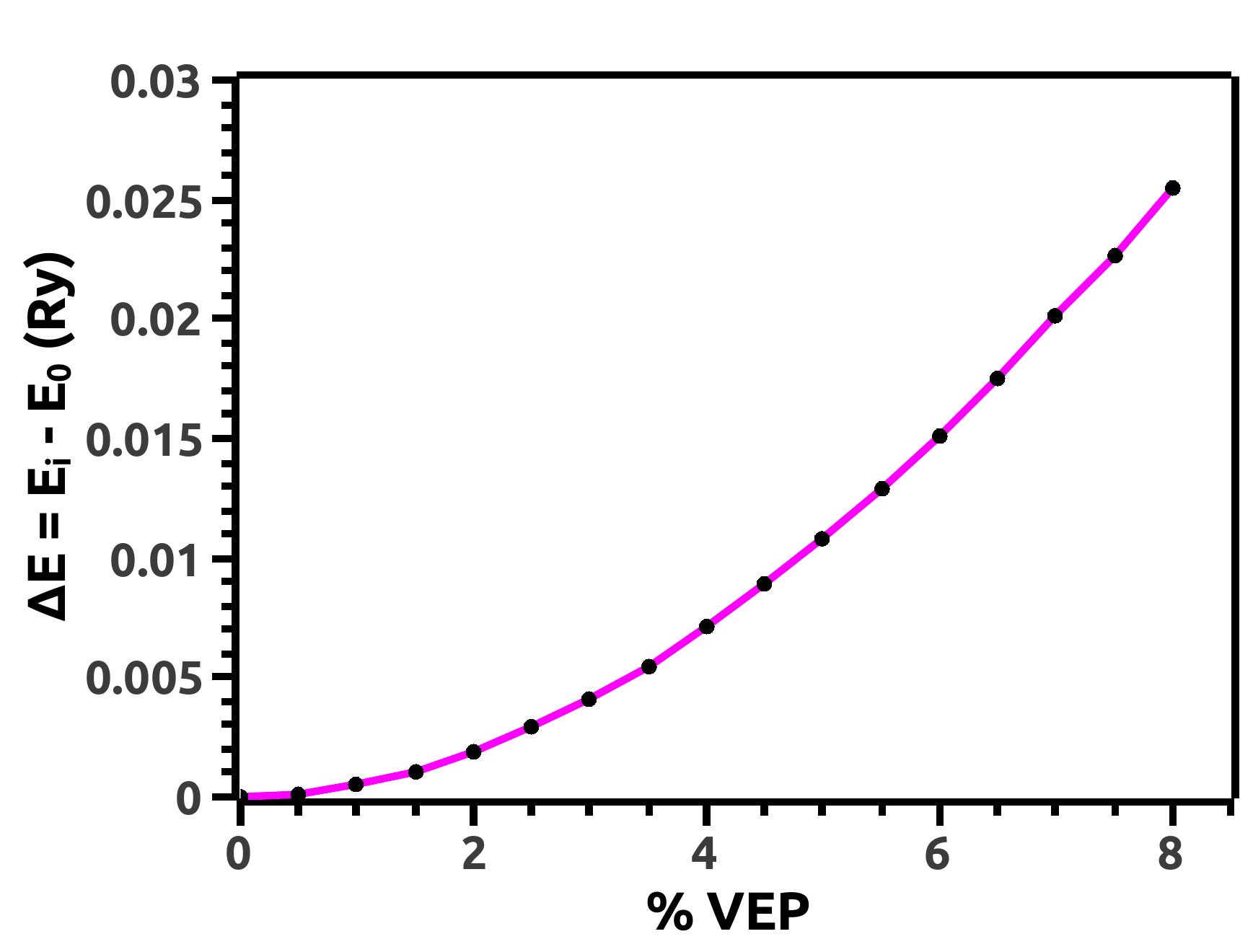}
	\caption{Change in Total Energy (Ry) of the system indicating a shift from the equilibrium state leading to a quantum phase transition.}
	\label{fig:delE}
\end{figure}

We increase lattice constant (a) in steps of 0.5\% from 0.0\% to 8.0\% to investigate for TPT in HH LiMgBi. We refer to this increment as VEP which can be attributed to an effect created by the presence of an intrinsic \textit{void} (exerting a volume isotropic pressure in the outward direction) in the crystal structure or due to extreme internal \textit{thermal} perturbations. Also, it can be attributed to experimentally observed phenomena such as \textit{cavitation} pressure \cite{herbert2006cavitation} due to impurities known as \textit{cavity nuclei}. Increment in unit cell volume (a.u.$^3$) due to application of VEP (Figure \ref{fig:unit_vol}) is interpreted in terms of volume strain as shown in Figure \ref{fig:vol_str}. Also, we analyse the difference in Total Energy ($\Delta$E = E$_i$ - E$_0$) of the system (in Ry units) with respect to \% VEP (Figure \ref{fig:delE}) due to the volume strain and find that, due to increment in VEP, the system deviates from its equilibrium state leading to a quantum phase transition.

\begin{figure}[h]
	\centering
	\includegraphics[width = 4in]{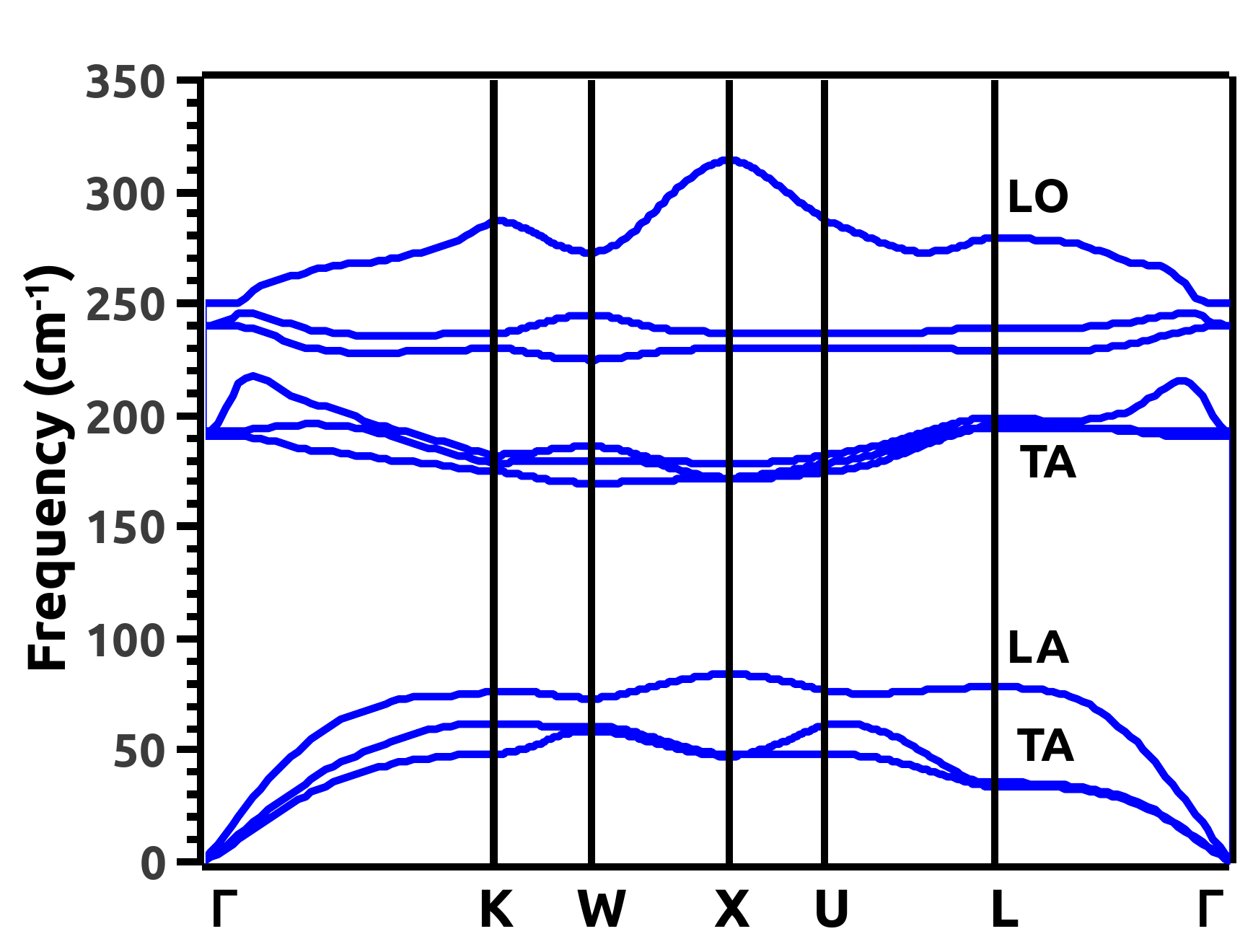}
	\caption{Phonon Dispersion Curves at 0\% VEP}
	\label{fig:phonon}
\end{figure}

Phonon dispersion calculations were performed to check the lattice dynamical stability of the proposed crystal system LiMgBi belonging to the F$\overline{4}$3m[216] space group. This is important because, Phonon studies are fundamental in underpinning the vibrational dynamics in a phase transition and the practical feasibility to synthesize the material \cite{pillai2018strain}. In the current study, Phonon Dispersion Curves (PDC) are obtained as shown in Figure \ref{fig:phonon} at 0\% VEP. As there are three (3) atoms in the unit cell of LiMgBi, we obtain nine (9) Phonon branches in the dispersion relation (Figure \ref{fig:phonon}). The PDC shows that, the system LiMgBi is, dynamically stable and has no imaginary modes of vibration. From Figure \ref{fig:phonon}, we can see that, in the lower frequency regime there are 3 modes which correspond to the acoustic branches and in the higher frequency regime, there are six (3N - 3, here N = 3) optical modes \cite{chen2005nanoscale}. The acoustic branch exhibits one (in plane) Longitudinal Acoustic (LA) and two (in plane) Transverse Acoustic (TA) modes of vibration which are degenerate along the directions, \textbf{W} to \textbf{X} and \textbf{L} to $\mathbf{\Gamma}$ in the BZ. The optical branch exhibits two (N - 1, here N = 3) Longitudinal Optical (LO) and four (2N - 2, here N = 3) Transverse Optical modes of vibration which dominate in the higher frequency regime \cite{chen2005nanoscale}. The TO modes are degenerate along the directions $\mathbf{\Gamma}$ to \textbf{K}, \textbf{K} to \textbf{X} and \textbf{X} to $\mathbf{\Gamma}$ in the BZ. Similarly, the LO modes are degenerate at $\mathbf{\Gamma}$ in the BZ.

With proper understanding of the structural and dynamic stability of LiMgBi, we proceed for further analysis. We shall broadly divide our discussions here off in two perspectives, \textit{qualitative} which revolves around the electronic properties and \textit{quantitative} which revolves around the $\mathbb{Z}_2$ analysis.

\subsubsection*{Electronic Properties and Surface States}

\begin{figure}[h]
	\centering
	\includegraphics[width = 4in]{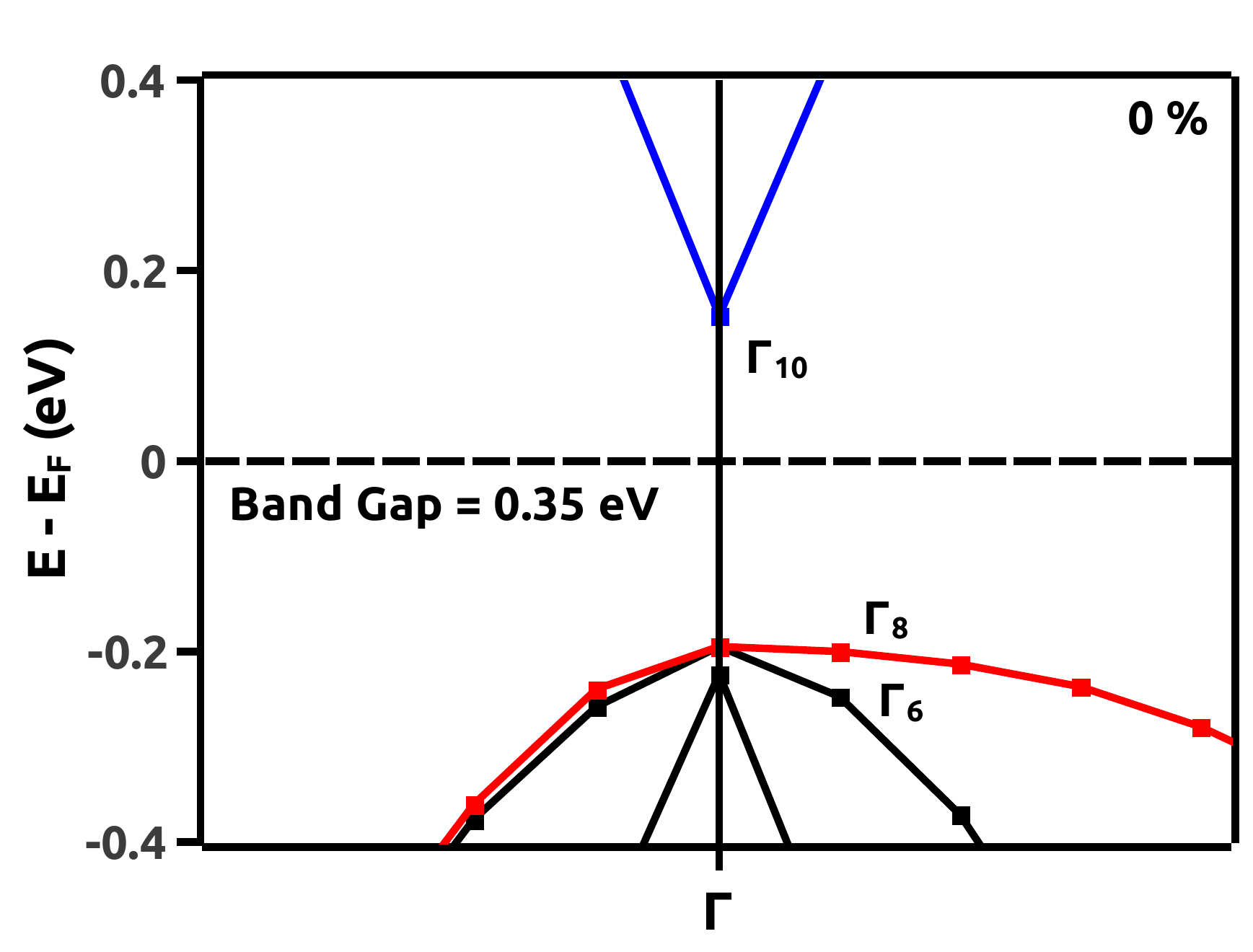}
	\centering
	\includegraphics[width = 4in]{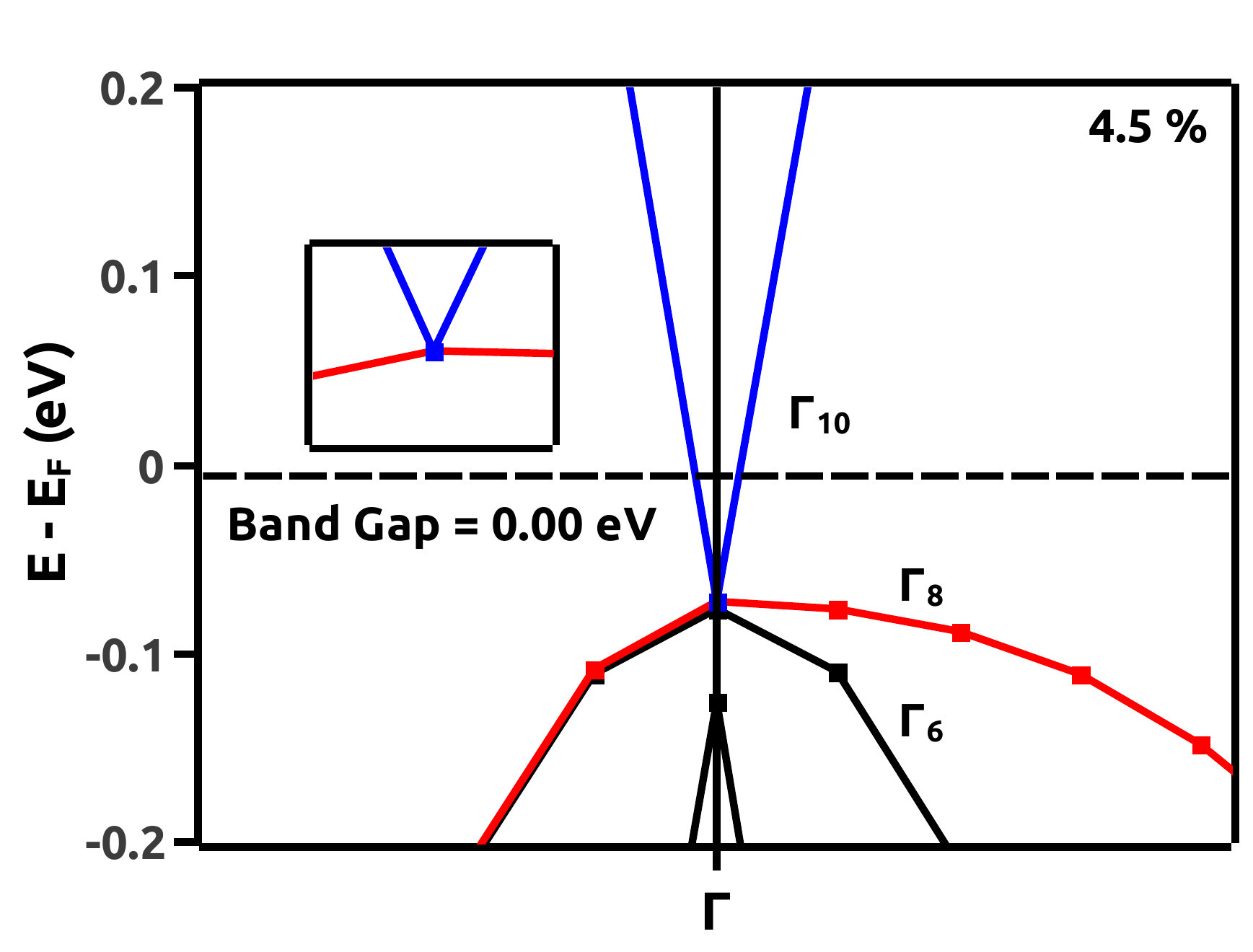}
	\centering
	\includegraphics[width = 4in]{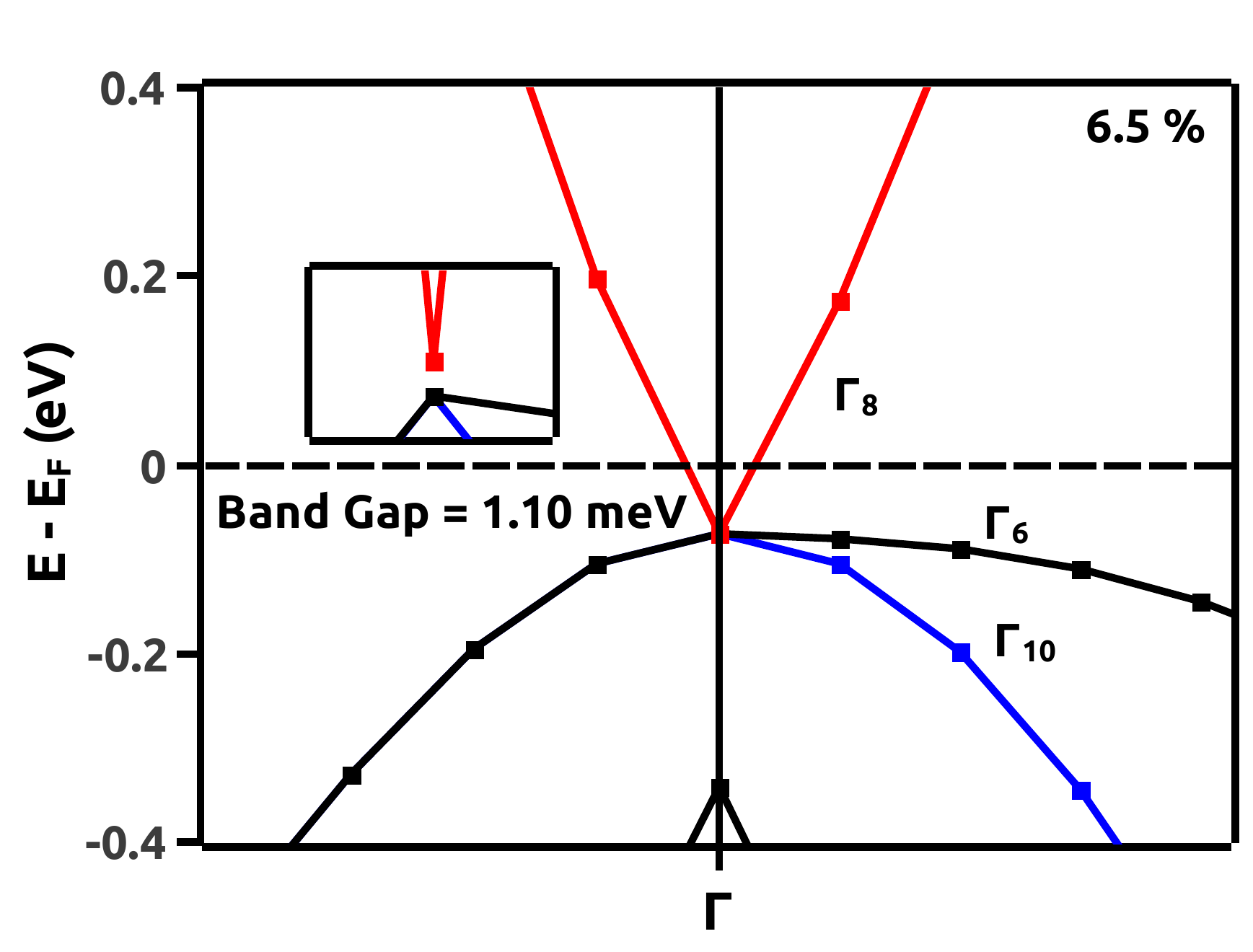}
	\caption{EBS at 0\%, 4.5\% and 6.5\% VEP}
	\label{fig:bands}
\end{figure}

EBS analysis has proved to be a fruitful \textit{qualitative} method to look for TI nature in materials \cite{chadov2010tunable,zhang2011band,ando2013topological,agapito2013novel,pal2014strain,aramberri2018strain,monserrat2017antiferroelectric}. We hence, analyse the EBS for band inversion by band engineering which is a characteristic signatures of TI nature. It is observed that, band inversion occurs in LiMgBi when subjected to VEP in steps of 0.5\% at a critical value of 4.0\%. From the EBS we observe a phase transition from a band insulator like behaviour to a Dirac semi-metal like behaviour. A direct BG (0.35 eV and 0.70 eV from PBE and HSE06 calculations respectively) exists at high symmetry point ($\mathbf{\Gamma}$) in the BZ at 0\% VEP (Figure \ref{fig:bands}). This is compared to previous studies (Table \ref{tab1}) where the band gap is $\sim$ 0.62 eV \cite{roy2012half}. BG from our HSE calculations match closely with the BG from previous studies \cite{roy2012half}. For qualitative analysis we continue to discuss the electronic properties in terms of band gap under GGA and HSE approximation (tabulated in Table \ref{tab2} for three different values of \% VEP). 

\begin{table}[ht]
	\centering
	\begin{tabular}{c c c}
		\hline
		Functional & \% VEP & Band Gap (eV) \\
		\hline\hline
		\multirow{3}{4em}{GGA} & 0.0 & 0.35 \\ 
		& 4.0 & 0.00 \\
		& 6.5 & 0.001 \\
		\hline\hline
		\multirow{3}{4em}{HSE} & 0.0 & 0.70 \\ 
		& 4.0 & 0.00 \\ 
		& 6.5 & 0.10 \\
		\hline
	\end{tabular}
	\caption{Energy band gap from GGA and HSE calculations at 0.0\%, 4.0\% (critical VEP) and 6.5\% VEP.}
	\label{tab2}
\end{table}

It is observed from the band structure that, the Valence Band (VB) maxima has a doubly degenerate state along the high symmetry point $\mathbf{\Gamma}$ whereas, the Conduction Band (CB) minima has single non-degenerate state along the high symmetry point $\mathbf{\Gamma}$. This feature is protected by the two-fold symmetry of the FCC structure. With the gradual increment of VEP, at 4.0\% VEP, LiMgBi undergoes a phase transition from band insulator nature to a Dirac semi-metal nature (Figure \ref{fig:bands}) with BG 0.00 eV. We represent the eigen states along the high symmetry line $\mathbf{\Gamma}$ as $\mathbf{\Gamma}_i$ where `\textit{i}' is the band index without inclusion of spin degeneracy. The EBS for 0\%, 4.0\% and 6.5\% are shown in Figure \ref{fig:bands}. At 0\% VEP the band order is $\mathbf{\Gamma}_6$ (dominated by \textit{p} orbital), $\mathbf{\Gamma}_8$ (dominated by \textit{p} orbital) and $\mathbf{\Gamma}_{10}$ (dominated by \textit{s} orbital). Here, $\mathbf{\Gamma}_6$ and $\mathbf{\Gamma}_8$ bands are degenerate as seen in Figure \ref{fig:bands}(a). With increment in VEP, at 4.0\% a Dirac Cone is formed as shown in Figure \ref{fig:bands}(b). This feature exhibits a linear dissipationless transport of Fermions along the surface of LiMgBi. Also, the formation of a Dirac Cone leads to degeneracy between $\mathbf{\Gamma}_8$ and $\mathbf{\Gamma}_{10}$ bands as seen in Figure \ref{fig:bands}(b). The formation of Dirac cone also, lifts up the degeneracy between bands $\mathbf{\Gamma}_6$ and $\mathbf{\Gamma}_8$. By further increment in VEP, BG reopens with to 1.1 meV (under GGA) as shown in Figure \ref{fig:bands}(c) and 0.10 eV (under HSE). This reopening characterizes the band inversion in terms of the inversion in band order with, $\mathbf{\Gamma}_8$ (dominated by \textit{p} orbital) being exchanged with $\mathbf{\Gamma}_{10}$ (dominated by \textit{s} orbital) wherein, $\mathbf{\Gamma}_{10}$ feature is transferred to $\mathbf{\Gamma}_6$ (dominated by \textit{p} orbital) band. Also, as the CB crosses Fermi level it indicates that, LiMgBi exhibits conduction of Fermions along the surface.

\begin{figure}[h]
	\centering
	\includegraphics[width = 4in]{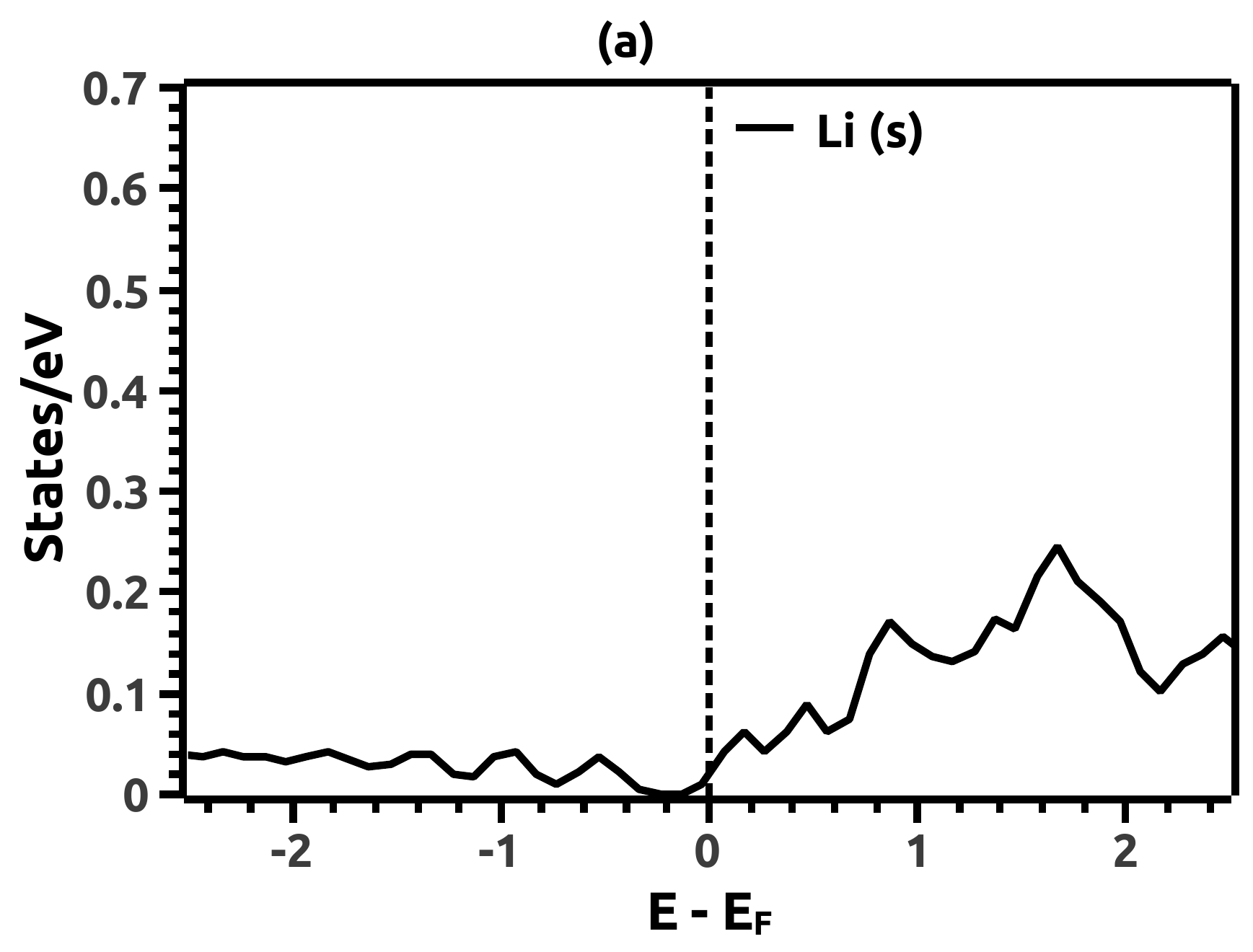}
	\centering
	\includegraphics[width = 4in]{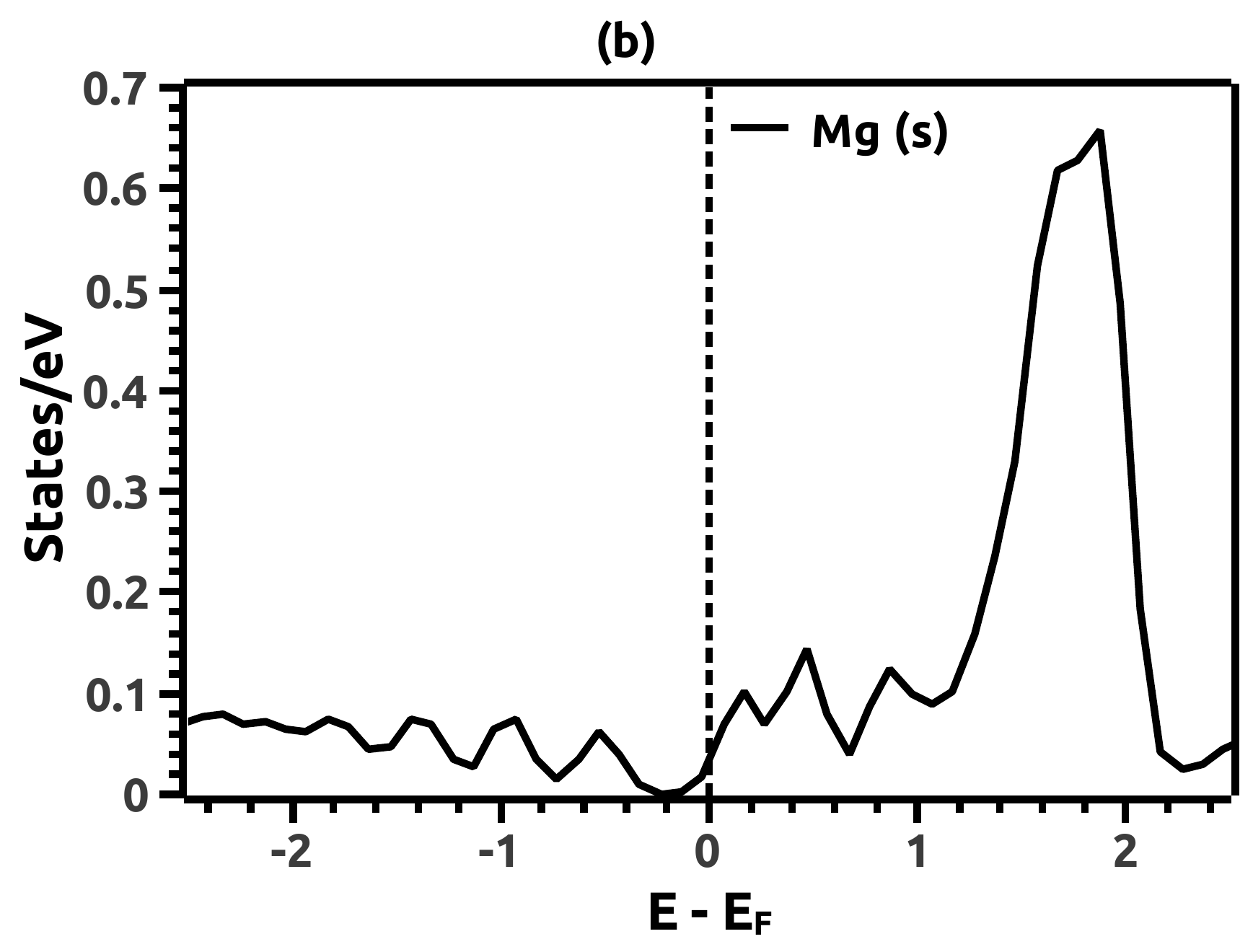}
	\centering
	\includegraphics[width = 4in]{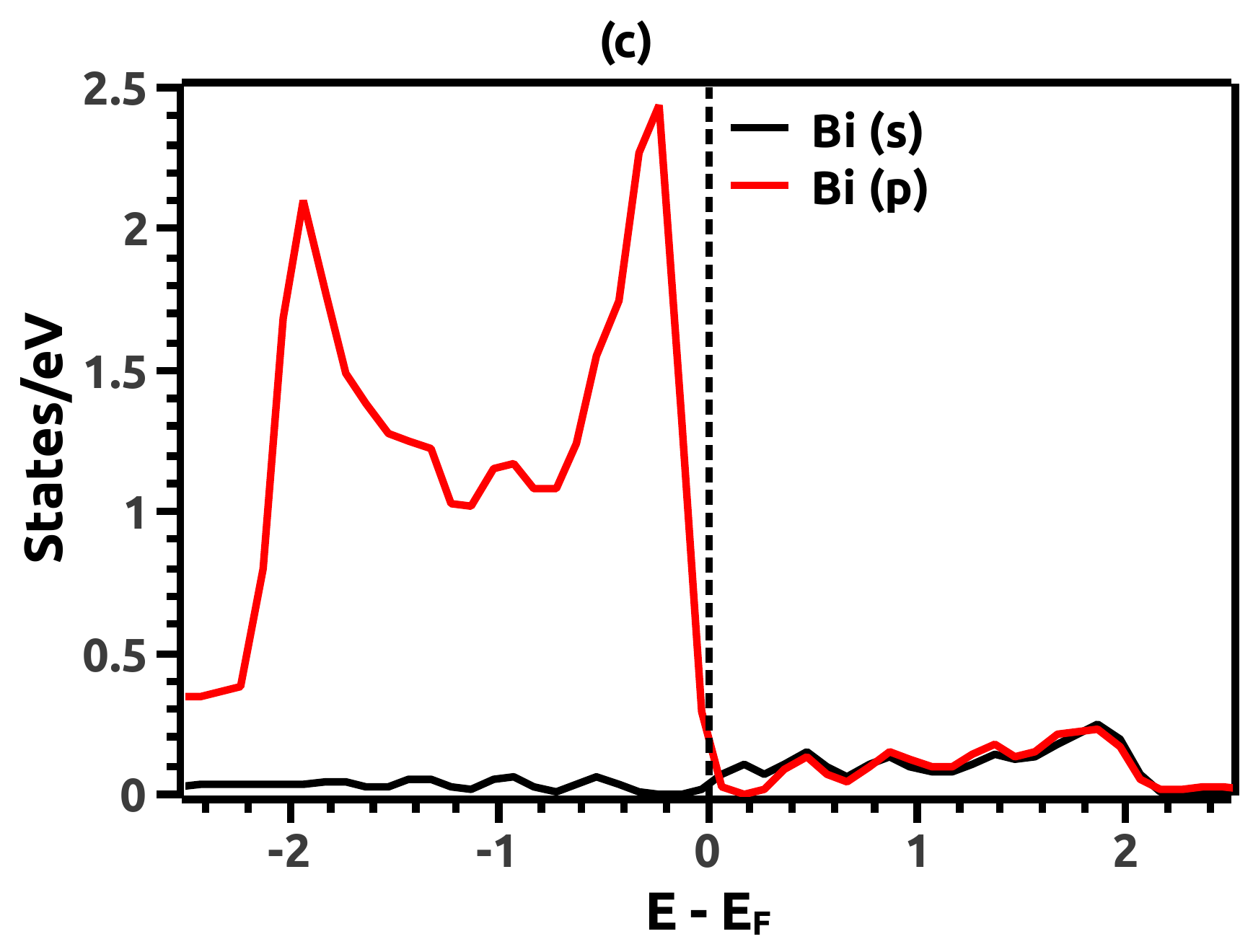}
	\caption{LDOS of (a) Li, (b) Mg and (c) Bi at 0\% VEP}
	\label{fig:ldos_0}
\end{figure}

\begin{figure}[h]
	\centering
	\includegraphics[width = 4in]{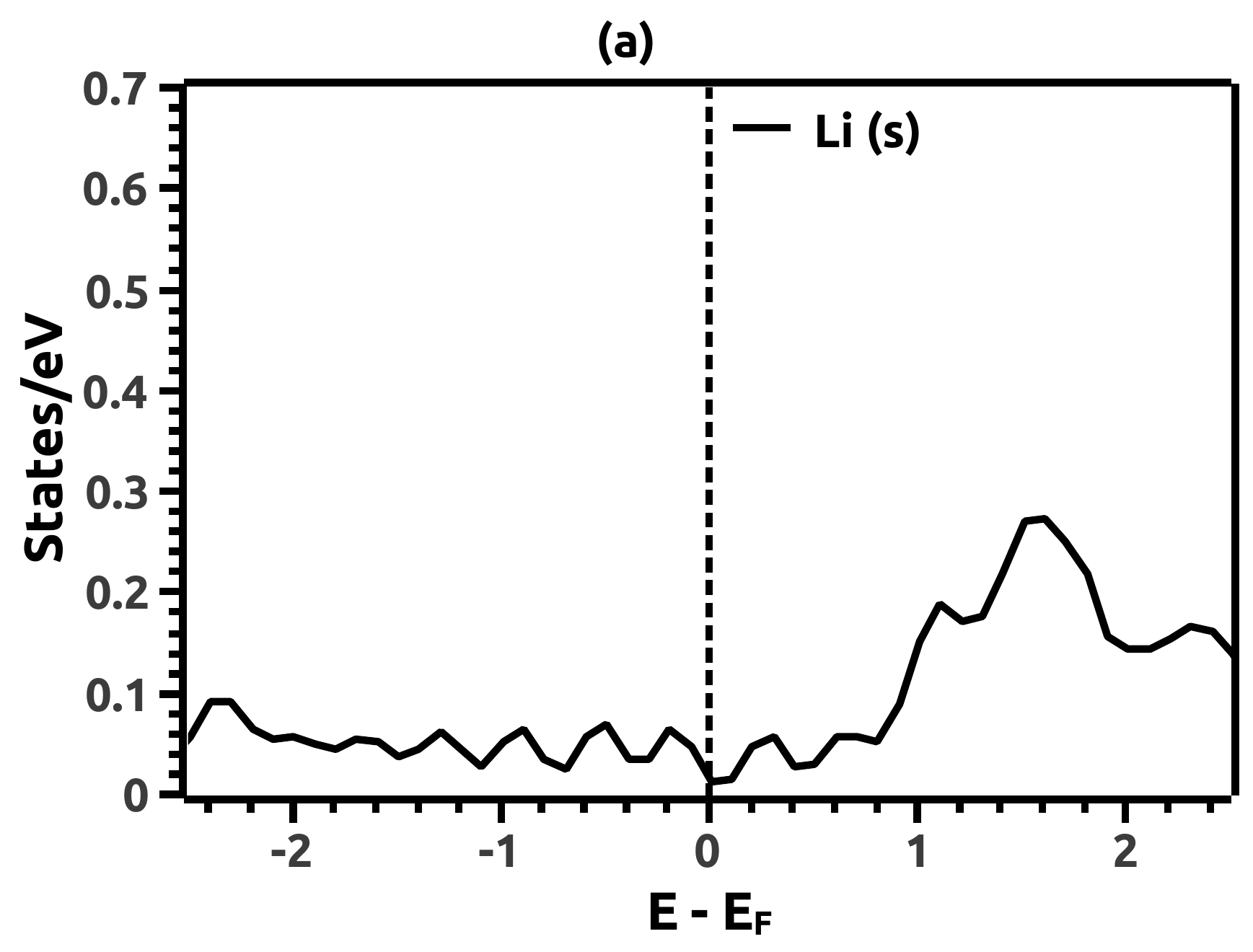}
	\centering
	\includegraphics[width = 4in]{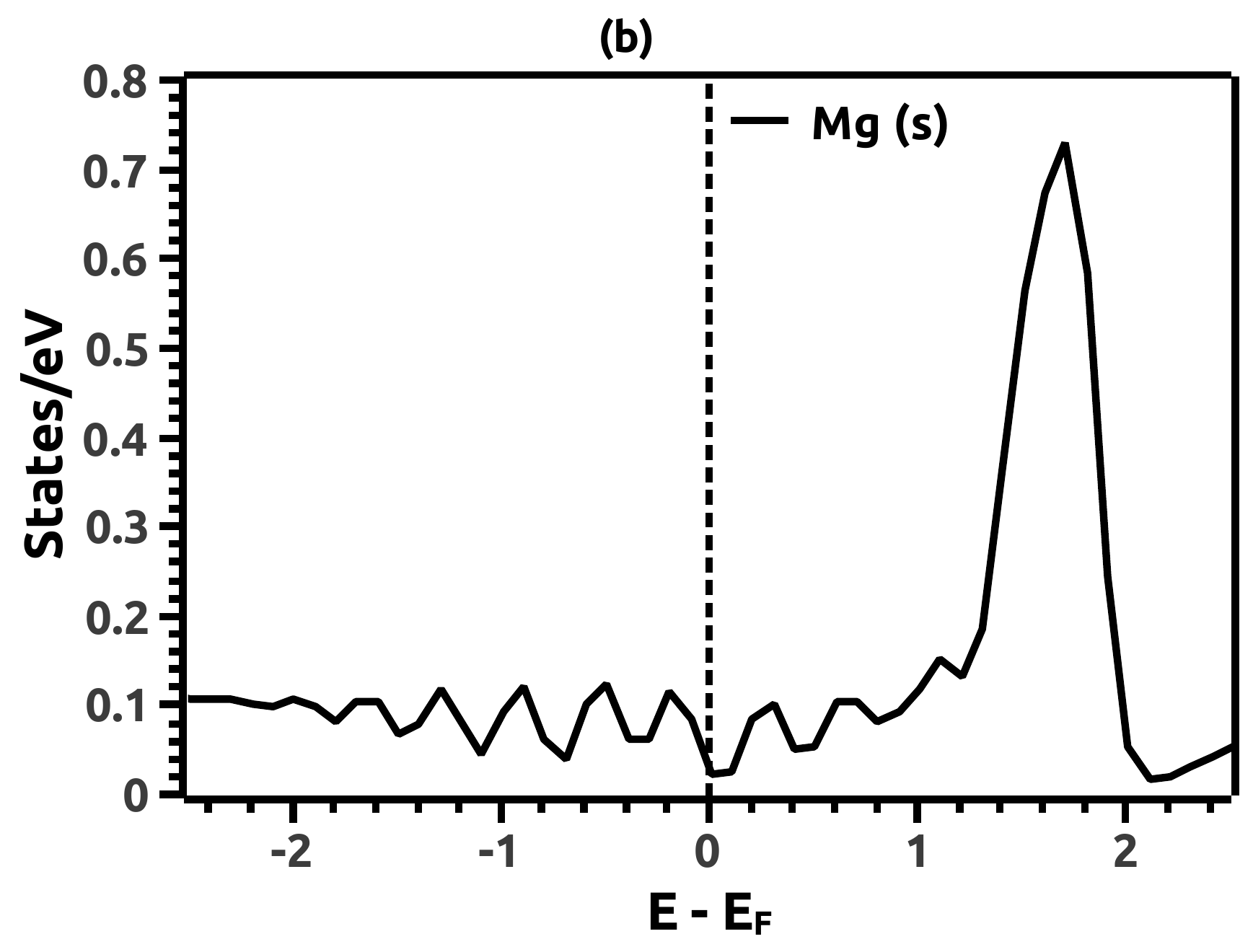}
	\centering
	\includegraphics[width = 4in]{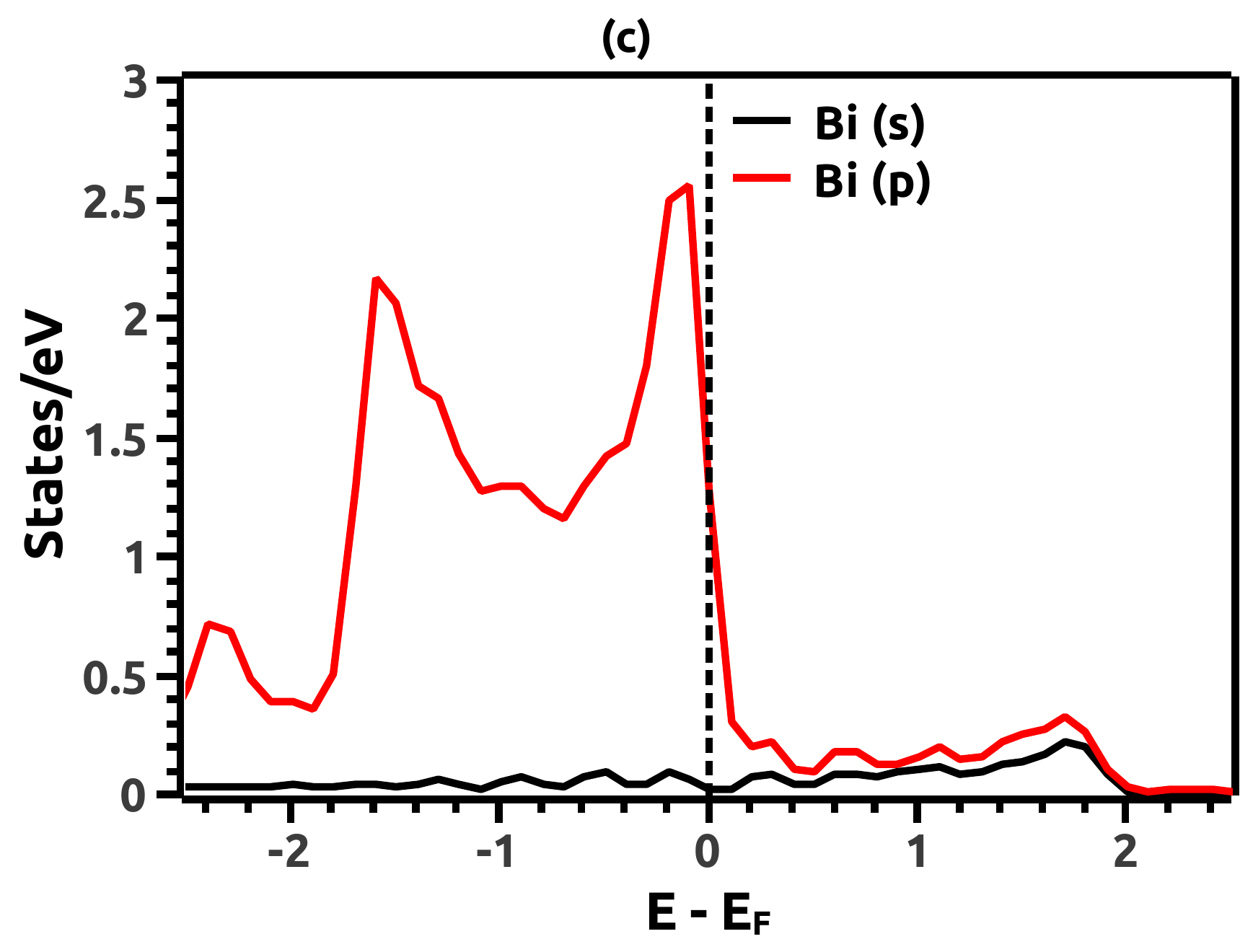}
	\caption{LDOS of (a) Li, (b) Mg and (c) Bi at 4.5\% VEP beyond the critical value}
	\label{fig:ldos_4_5}
\end{figure}

\begin{figure}[h]
	\centering
	\includegraphics[width = 4in]{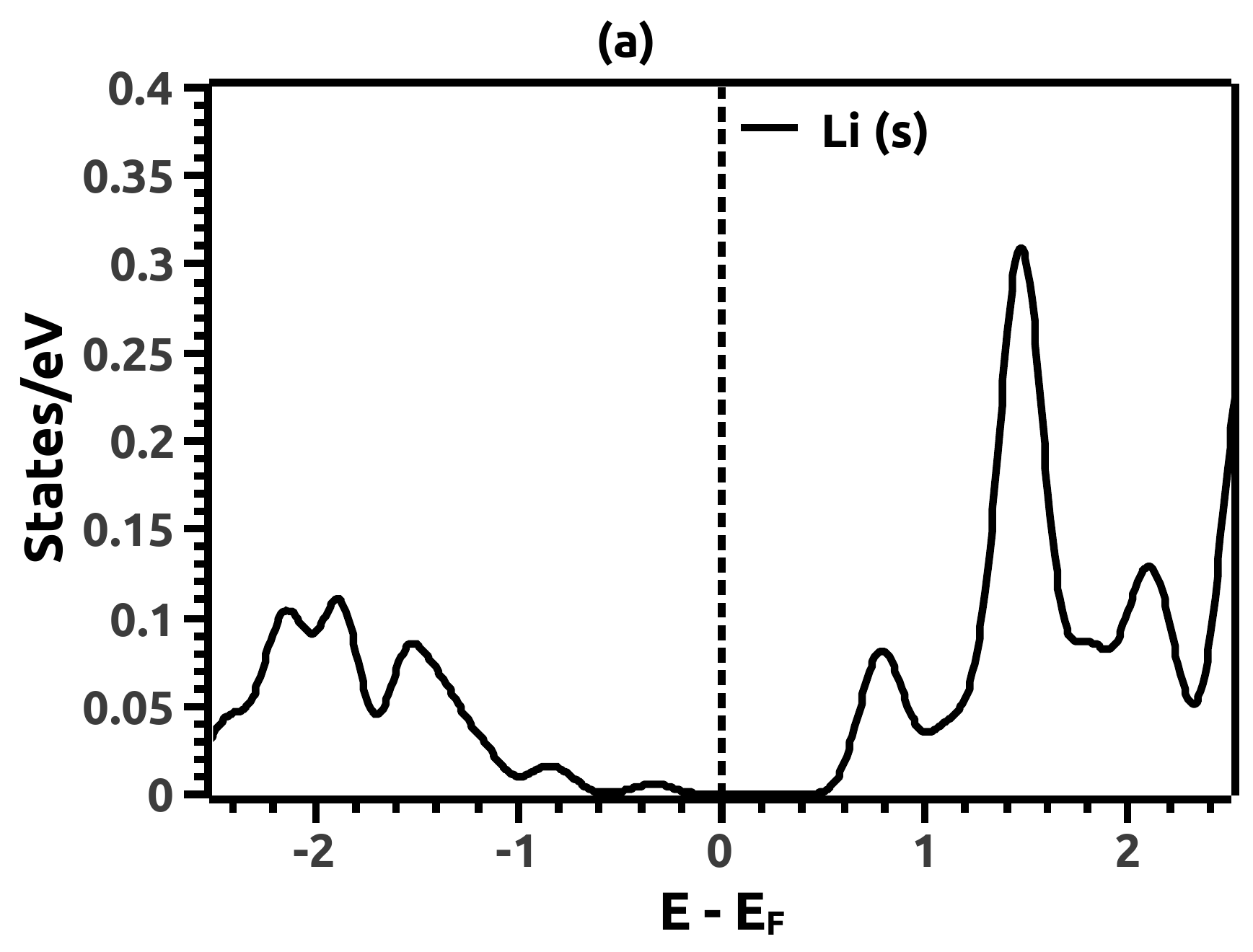}
	\centering
	\includegraphics[width = 4in]{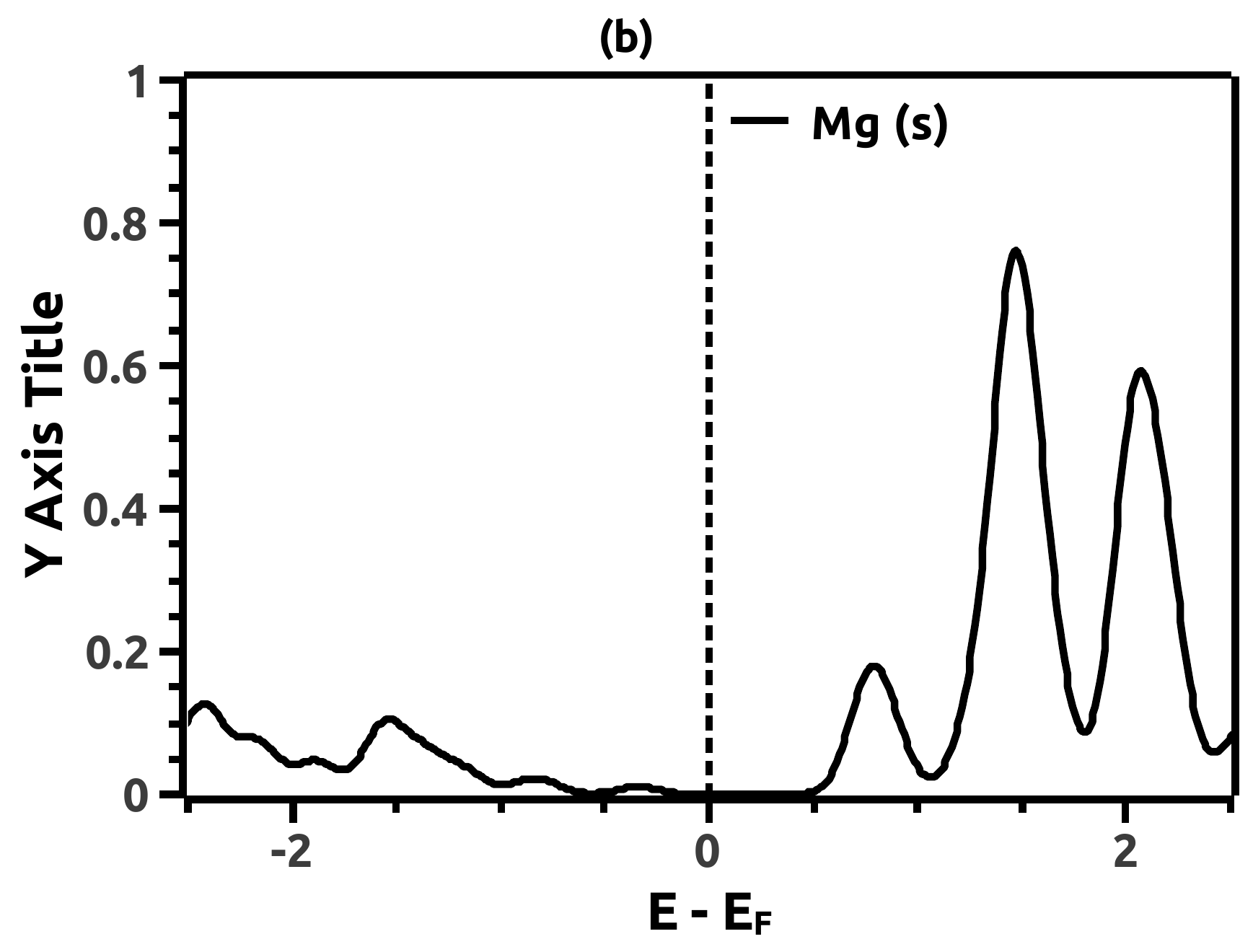}
	\centering
	\includegraphics[width = 4in]{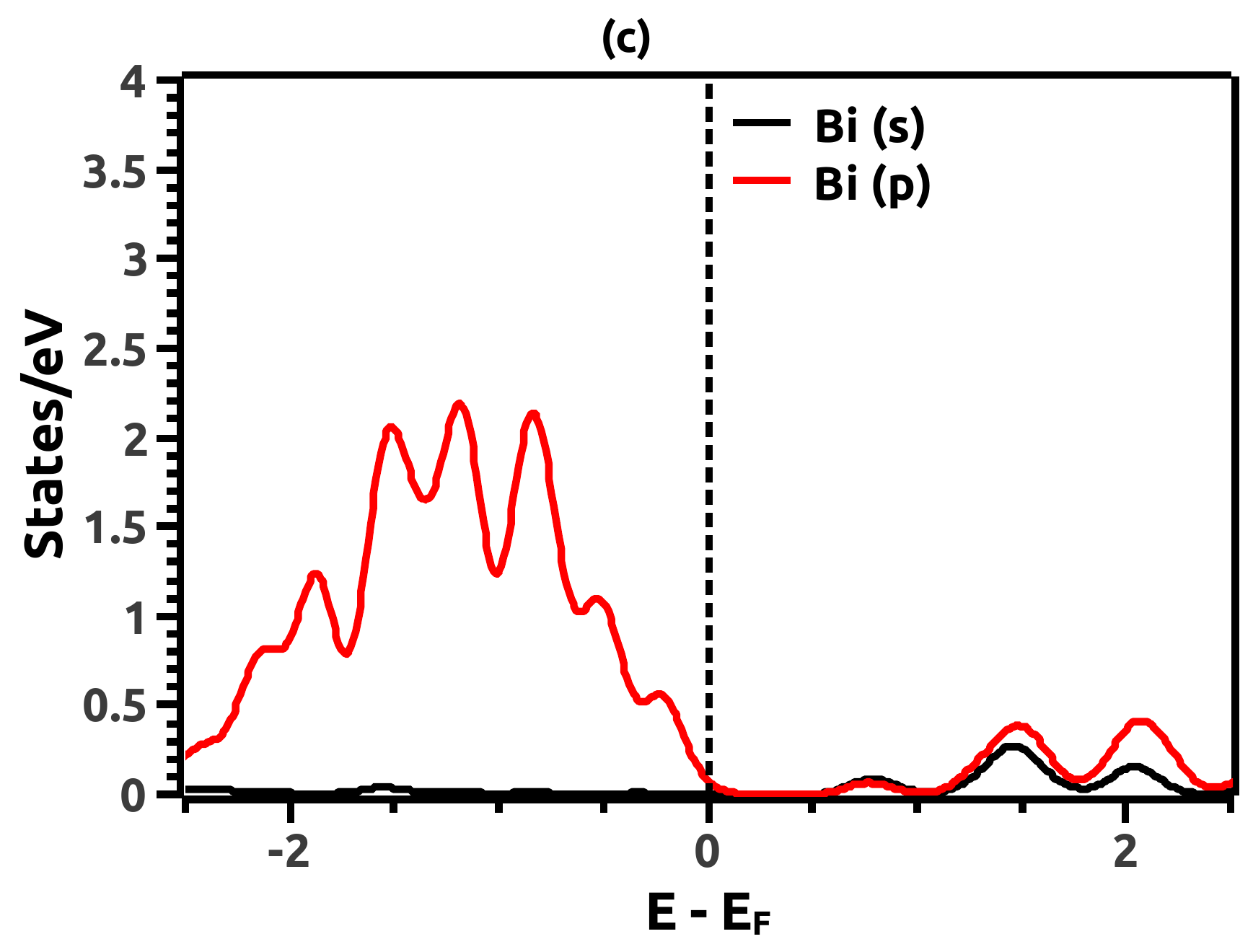}
	\caption{LDOS of (a) Li, (b) Mg and (c) Bi at 6.5\% VEP indicating band inversion}
	\label{fig:ldos_6_5}
\end{figure}

In order to justify the band inversion, we discuss the orbital contributions of Li, Mg and Bi. For this purpose, projected LDOS were calculated to compliment the EBS. We refrain our discussions to the regions in the vicinity of Fermi energy level (E$_F$) which is important for interpretation of band inversion. At 0\% VEP, the LDOS is as shown in Figure \ref{fig:ldos_0}. VB in the vicinity of Fermi level in Li is dominated by a small contribution from unhybridized `s' orbital, while the CB is dominated majorly by unhybridized `s' orbital. Similarly, the VB in Mg has small contributions from the `s' orbitals and the CB has major contribution from the `s' orbital, both being unhybridized. In both the cases (Li and Mg), the `s' orbital contribution from the CB crosses Fermi energy level (E$_F$) as seen from Figure \ref{fig:ldos_0}.  As compared to this, Bi has a strong contribution solely due to the `p' orbital in VB with a strong hybridization with the `s' orbital in CB. Also, in the VB a mild contribution from the `s' orbital is observed but, near Fermi energy level (E$_F$) the `s' orbital has major contribution as seen in Figure \ref{fig:ldos_0}. Beyond a critical value of VEP at 4.5\% the orbital contributions for the Dirac cone formation at the Dirac point on the Fermi energy level (E$_F$) are interpreted from the LDOS as shown in Figure \ref{fig:ldos_4_5}. The VB and CB of Li and Mg are dominated by the `s' orbital with a increment in its orbital contribution at E$_F$ indicating the formation of Dirac cone (Figure \ref{fig:ldos_4_5}). Although, the CB in both (Li and Mg) retains the `s' orbital dominant feature. Similarly, in Bi, the major contribution for Dirac cone formation comes from the VB `p' orbital crossing the Fermi energy level (E$_F$) which has a weak hybridization with `s' orbital, the strong hybridization persists and dominates the deeper regions of CB (Figure \ref{fig:ldos_4_5}). At 6.5\% of VEP when the band reopens, the orbital contributions change as compared to the 0\% and 4.5\% case above. Figure \ref{fig:ldos_6_5} shows the LDOS at 6.5\% VEP. From the LDOS for Li and Mg, we observe that, the orbital contribution due to `s' orbital increases in the VB region (Figure \ref{fig:ldos_6_5}). Similarly from the LDOS for Bi, the `p' orbital contribution reduces in the VB and increases in the interior regions of CB with a strong hybridization with the `s' orbitals, whereas, the `s' orbital contribution is almost negligible in VB but increases slightly in the interior of VB. The changes observed in the orbital contributions are contemplated as an indicator of the band order rearrangement observed in the EBS, which is a characteristic signature of band inversion. The major contribution for the band inversion is due to the Bi which has strong share in orbital rearrangements.

\begin{figure}[h]
	\centering
	\includegraphics[width = 4in]{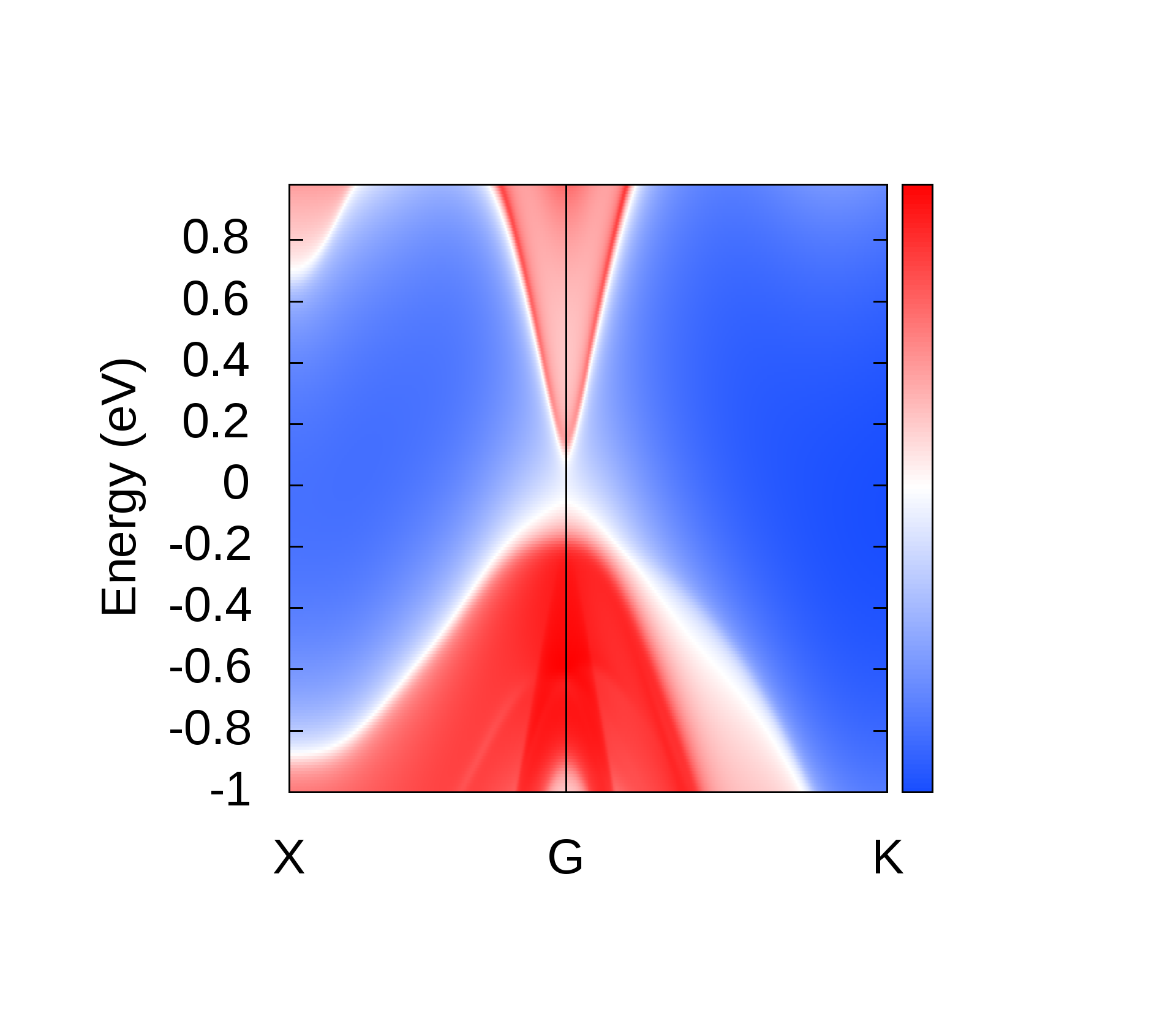}
	\centering
	\includegraphics[width = 4in]{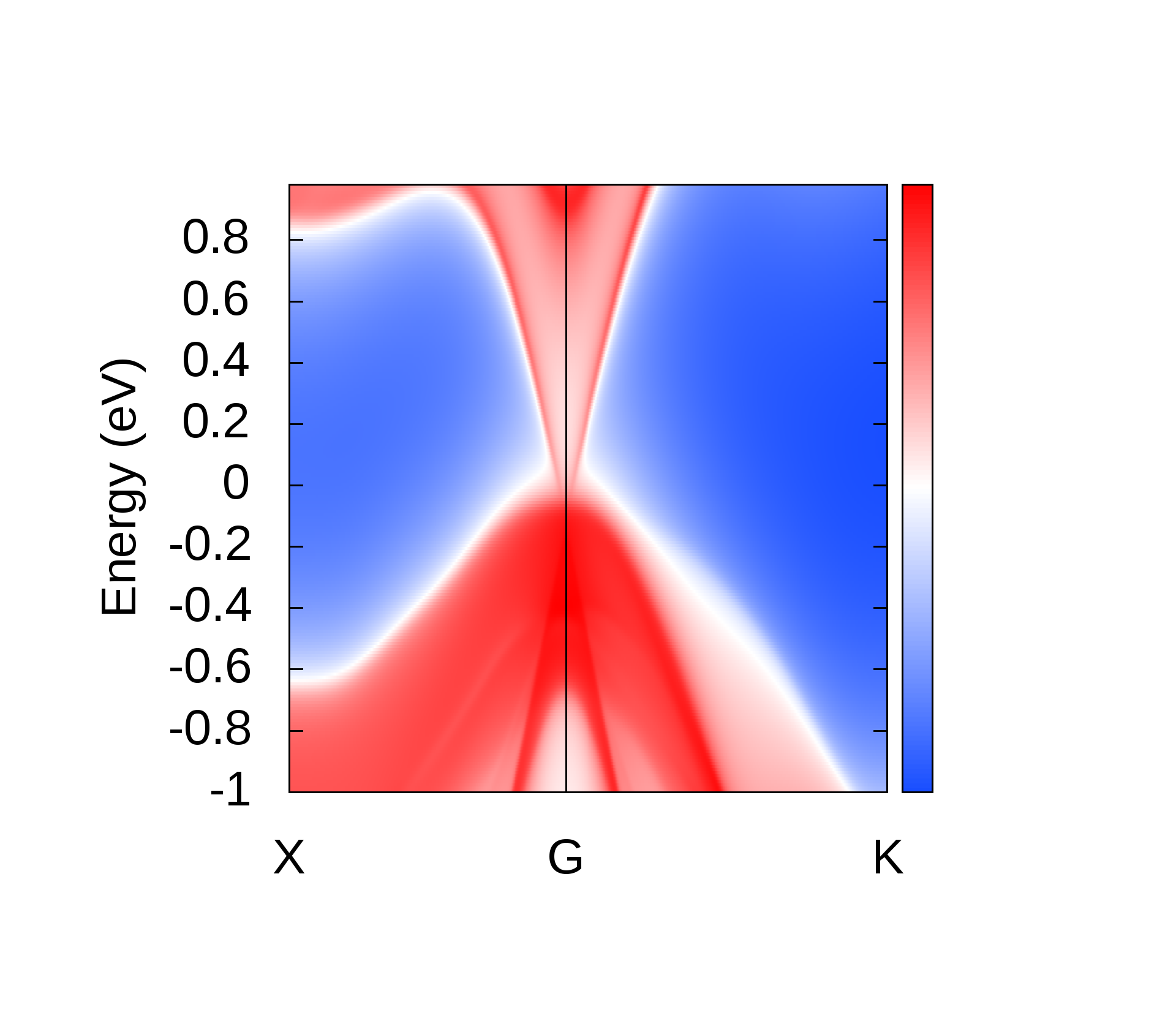}
	\caption{Surface States (computational ARPES) at 0\% and 4.5\% VEP obtained using WT}
	\label{fig:ss}
\end{figure}

Further analysis for a deeper realization of TI behaviour was performed by calculating the SS. Figure \ref{fig:ss} is the computational Angle Resolved Photo Emission Spectra (ARPES) \cite{chen2009experimental,hsieh2009observation,ando2013topological} showing the SS \cite{PhysRevB.83.205101} at 0\% and 4.5\% VEP respectively. Beyond the critical value of VEP (i.e., beyond 4.0\% VEP) we observe dissipationless Fermi Surface along the edge states in the BZ. This is a qualitative evidence of TI nature characterized by, a \textit{single} Dirac cone SS (similar to previous work \cite{xia2009observation}). This predicts that, experimental realization of FCC LiMgBi can be characterized by performing ARPES to examine the SS along the (0001) surface.

\subsubsection*{$\mathbb{Z}_2$ Analysis}

\begin{figure}[h]
	\centering
	\includegraphics[width = 4in]{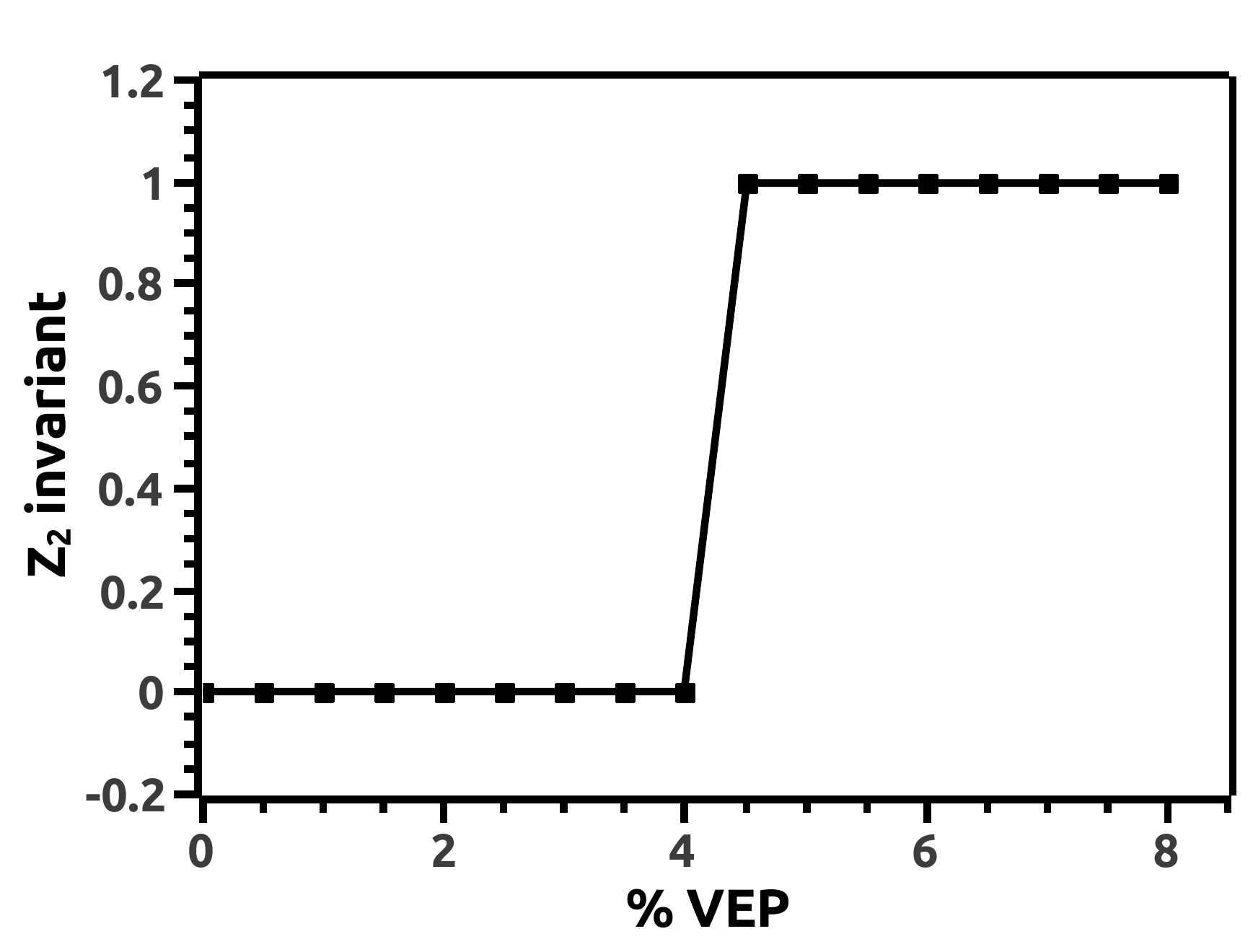}
	\caption{$\mathbb{Z}_2$ evolution along the applied \% VEP}
	\label{fig:z2}
\end{figure}

The $\mathbb{Z}_2$ invariant is essential to characterize a material into a $\mathbb{Z}_2$ topological class. There are various methods to obtain the $\mathbb{Z}_2$ invariant \cite{gresch2017z2pack,soluyanov2011computing,doi:10.1143/JPSJ.76.053702,yu2011equivalent,fu2007topological} for systems with and without inversion symmetry. Generally, for systems with inversion symmetry, the product of parities of eigen values is calculated at eight (8) Time Reversal Invariant Momenta (TRIM) points for a 3D system (equation \ref{prod}) and at four (4) TRIM points for a two dimensional (2D) system \cite{fu2007topological}. The $\mathbb{Z}_2$ invariant $\nu_0 = 1$ and $\nu_0 = 0$ corresponds to strong and weak TI respectively \cite{fu2007topological,ando2013topological}. Evolution of $\mathbb{Z}_2$ invariant along the application of pressure is shown in Figure \ref{fig:z2}. Since, the system is FCC, which lacks inversion symmetry, the $\mathbb{Z}_2$ analysis was performed by calculating the $\mathbb{Z}_2$ invariants $\big( \mathbb{Z}_2 \big)_\pi$ and $\big( \mathbb{Z}_2 \big)_0$, for two different planes of momentum (\textit{k}) in the BZ resulting in the $\mathbb{Z}_2$ invariant $\nu_0$ as, $\nu_0 = \Big[ \big( \mathbb{Z}_2 \big)_\pi - \big( \mathbb{Z}_2 \big)_0 \Big] \big( mod 2 \big)$. These calculations were performed using the WannierTools (WT) package \cite{WU2017} which utilizes the information regarding the Wannier Function Centres (WFC) obtained by post processing package Wannier90 (W90) \cite{MOSTOFI20142309}. Maximally Localised Wannier Functions (MLWF) were obtained after PWSCF calculations using Marzari and Vanderbilt (MV) method in W90 package. The MV method minimizes the gauge dependent spread $\tilde{\Omega}$ with respect to the Bloch states U$^{(k)}$ which are obtained from the PWSCF calculations \cite{QE-2017}. The WFC are then used for computation of Wilson Loops in WT. WT was used to obtain Tight Binding (TB) model for LiMgBi using the MLWF from W90. Using this TB model, we compute the $\mathbb{Z}_2$ invariant by calculating the Wilson loops around the Wannier Charge Centres (WCC). This method is one of the best methods for systems without inversion symmetry \cite{WU2017}. Computational APRES is obtained using WT along the \big(0001\big) crystal direction. WT is a reliable software used quite often for such studies \cite{soluyanov2011computing,yu2015topological,weng2015topological,li2016dirac,bzduvsek2016nodal}.

\begin{equation}
(-1)^{\nu_i} = \prod_{i = 1}^{8} \delta(\Lambda_{i})
\label{prod}
\end{equation}

\begin{equation}
\nu_0 = \Big(\mathbb{Z}_2 \big(k_i = 0 \big) + \mathbb{Z}_2 \big(k_i = 0.5 \big) \Big) mod 2
\label{sum}
\end{equation}

\begin{equation}
\nu_i = \mathbb{Z}_2 \big(k_i = 0.5 \big)
\label{sum2}
\end{equation}

The underlying computation in WT is carried out along the six Time Reversal Invariant Planes (TRIP) k$_x$ = 0,$\pi$, k$_y$ = 0,$\pi$ and k$_z$ = 0,$\pi$ in the BZ. The $\mathbb{Z}_2$ indices ($\nu_0$, $\nu_1$ $\nu_2$ $\nu_3$) are computed using equation \ref{sum} and \ref{sum2} \cite{WU2017}. Based on these calculations, LiMgBi under 0\% VEP is characterized by $\mathbb{Z}_2$ invariant $\nu_0$ = 0 which implies that LiMgBi is a weak TI. But, beyond the critical value of VEP, at 4.5\% the $\mathbb{Z}_2$ indices are obtained to be \big($\nu_0$, $\nu_1$ $\nu_2$ $\nu_3$\big) $\equiv$ \big(1, 0 0 0\big) which implies that, LiMgBi is a strong TI. Thus we conclude quantitatively that, LiMgBi undergoes a TPT under the application of VEP.

\subsection*{Conclusion}

In summary, we performed density functional theory based \textit{first-principles} calculations and find that, HH compound LiMgBi which was previously studied for its semi-conducting, piezoelectric and thermo-electric properties undergoes a TPT driven by VEP. The FCC structure of LiMgBi was subjected to VEP isotropically by increment of the lattice constant (a). At equilibrium, the EBS has a direct BG of $\sim$ 0.35 eV (under GGA) and $\sim$ 0.70 eV (under HSE06). With gradual increment in VEP, we find a critical value at which, LiMgBi undergoes a \textit{phase transition} from a trivial band insulator to a Dirac semi-metal material. With further increment in VEP, the band reopening is observed which is quite small; with a BG of the order of $\sim$ 1.1 meV (under GGA) and $\sim$ 0.10 eV (under HSE) indicating, band inversion due to exchange of orbital contribution in the VBM and CBM. This was complemented by thorough analysis of projected LDOS in order to understand the orbital contribution of the constituent elements. In order to indicate experimental characterization of the TI nature, SS were calculated using computational ARPES along the direction \big(0001\big) which exhibits the presence of a surface Dirac cone. These results were further quantified by studying the evolution of $\mathbb{Z}_2$ invariant along the application VEP. The strong TI nature of HH LiMgBi is characterized by the $\mathbb{Z}_2$ index, \big($\nu_0$, $\nu_1$ $\nu_2$ $\nu_3$\big) $\equiv$ \big(1, 0 0 0\big). We thus conclude that, HH compound LiMgBi with FCC crystal structure undergoes a TPT driven by VEP. These calculations will open up diverse perspectives for multi-purpose applications of LiMgBi as a strong TI apart from the known semi-conducting, piezoelectric and thermo-electric applications.

\subsection*{Acknowledgement}
Authors acknowledge, Department of Science and Technology (DST), Government of India, for providing financial assistance.


\pagebreak
\bibliography{LiMgBi}
\bibliographystyle{unsrt}

\pagebreak
\section*{Figures}

\end{document}